\pdfoutput=1
\documentclass{JINST}

\title{Performance of the Hamamatsu R11410 Photomultiplier Tube in Cryogenic Xenon Environments}

\author{Laura Baudis$^a$, Annika Behrens$^a$, Alfredo Ferella$^a$, Alexander Kish$^a$, Teresa~Marrod\'{a}n~Undagoitia$^{a,b}$, Daniel Mayani$^a$, Marc Schumann$^{a,c}$\thanks{Corresponding
author.} \\
\llap{$^a$}Physik Institut, Universit\"{a}t Z\"{u}rich, 8057 Z\"{u}rich, Switzerland\\
\llap{$^b$}Max-Planck-Institut f\"{u}r Kernphysik, 69117 Heidelberg, Germany\\
\llap{$^c$}Albert Einstein Center for Fundamental Physics, Universit\"{a}t Bern, 3012 Bern, Switzerland\\
  E-mail: \email{marc.schumann@lhep.unibe.ch}}

\abstract{The Hamamatsu R11410 photomultiplier, a tube of 3``~diameter and with a very low intrinsic radioactivity, is an interesting light sensor candidate for future experiments using liquid xenon (LXe) as target for direct dark matter searches. We have performed several experiments with the R11410 with the goal of testing its performance in environments similar to a dark matter detector setup. In particular, we examined its long-term behavior and stability in LXe and its response in various electric field configurations.  }


\begin{document}

\section{Introduction}

Numerous astronomical and cosmological observations indicate that the majority of the matter content of our Universe consists of a yet unknown form of dark matter~\cite{ref::dm}. A well motivated candidate which arises naturally in many extensions of the standard model of particle physics is a weakly interacting massive particle (WIMP)~\cite{ref::wimp} which might be detected in ultra-low background detectors with energy thresholds around a few~keV or below~\cite{ref::directdet}. In order to protect these from the background caused by cosmic rays they are located in underground laboratories and are further placed inside passive shields against radioactivity from the surroundings. The signature from elastic scattering of WIMPs off a target nucleus is a steeply falling nuclear recoil spectrum with a maximum recoil energy of a few tens of~keV.

There are various experimental attempts to directly detect WIMP dark matter~\cite{ref::dmrev}. Experiments using the noble gas xenon in its liquid form are currently the most sensitive ones. In particular XENON100~\cite{ref::xeInstr}, located at the Gran Sasso Laboratory (LNGS)~\cite{ref::LNGS} in Italy, has recently published the most sensitive upper limits on WIMP-nucleon cross-sections over a large range of WIMP masses~\cite{ref::xe100_run10,ref::xe100_run10sd}. Liquid xenon detectors are often operated as dual-phase time projection chambers (TPC), where the scintillation light emitted by the xenon after a particle interaction, as well as the liberated ionization charges are measured simultaneously (see~\cite{ref::xeInstr} and references therein). The knowledge of both signals for every event allows reconstructing the interaction vertex in 3~dimensions and hence background suppression by selecting the inner part of the TPC via fiducialization, and for effective signal-background discrimination on an event-by-event basis.

The central part of these detectors are two arrays of photosensors, installed below the target volume in the liquid xenon (LXe) and above in the gas phase. Typical operation parameters are temperatures around $-100^\circ$C and absolute pressures of about 2\,bar. These sensors detect the primary scintillation signal, leading to typically only a few photoelectrons at threshold, as well as the charge signal which is amplified by proportional scintillation in the gas phase and therefore larger~\cite{ref::propScint}. For the current phase of experiments, photomultipliers (PMTs) are used because of their high quantum efficiency (QE), single photoelectron sensitivity, fast timing characteristics, low dark current, and because they can be operated at cryogenic conditions in LXe. The scintillation wavelength of xenon is in the vacuum ultraviolet region (VUV) at 178\,nm and the photocathode must have a high quantum efficiency at this wavelength. Preferentially it should be above 30\%, close to the maximum values achieved to-date.

In spite of ongoing efforts, for example within the DARWIN project~\cite{ref::darwin}, to replace photomultipliers by new types of sensors, such as hybrid photodetectors (QUPID~\cite{ref::qupid}) or gas photomultipliers (GPMs~\cite{ref::gpm}), PMTs still will be used in the ton-scale LXe dark matter detectors, as for example XENON1T~\cite{ref::xe1t}, LZ~\cite{ref::lz}, and Panda-X~\cite{ref::panda}. 

This article describes the Hamamatsu R11410~\cite{ref::hamamatsu}, a 3``~diameter PMT optimized for operation in LXe, which will be used in XENON1T. Its general performance is covered in~\cite{ref::lung2012} and its intrinsic radioactivity was evaluated in~\cite{ref::akerib2012}. Here, we focus on the most important aspect of the R11410 as candidate for large dark matter experiments, namely its performance at cryogenic temperatures immersed in LXe (as on the bottom PMT array of a TPC), in gaseous xenon (as on the top PMT array), and in electric fields (as in a TPC).

In Section~\ref{sec::setup}, we introduce the experimental setups used in the work presented here. Some general features of the PMT are summarized in Section~\ref{sec::character}. Results regarding the intrinsic radioactive contamination of this tube, a crucial parameter of any photosensor used for direct dark matter searches, are summarized in Section~\ref{sec::radioactivity}. The major part of this study, the long-term performance of the R11410 in LXe and the operation of a small test array in various high voltage configurations are presented in Sections~\ref{sec::lxestability} and \ref{sec::hvstability}, respectively. In Section~\ref{sec::conclusion} we summarize the main findings and give an outlook.

\section{Experimental Setup}\label{sec::setup}

In this Section, we briefly describe the various experimental setups operated at the University of Zurich used to characterize the R11410 presented in this study.

\paragraph{Data Acquisition and Gain Analysis} To measure the electron amplification gain of the PMTs, the emission of single photoelectrons (SPE) is stimulated by illumination of the photocathode with blue light ($\lambda = 470$\,nm) from an LED. The position of the peak in the SPE spectrum is proportional to the number of detected electrons and hence to the gain of the tube. A pulse generator (Telemeter TG4001) is used to bias the LED and to simultaneously trigger the data acquisition. The voltage is chosen such that an SPE pulse is seen on only about 3-5\% of the cases in order to suppress the contribution from two photoelectrons. The data presented here is acquired using a CAEN V1724 waveform digitizer with 100\,MHz sampling frequency and 40\,MHz input bandwidth. 
The waveforms are transferred to a computer and stored for data processing and subsequent analysis, and for later visual inspection. 

\begin{figure}[tb]
\centering
\includegraphics[width=0.76\columnwidth]{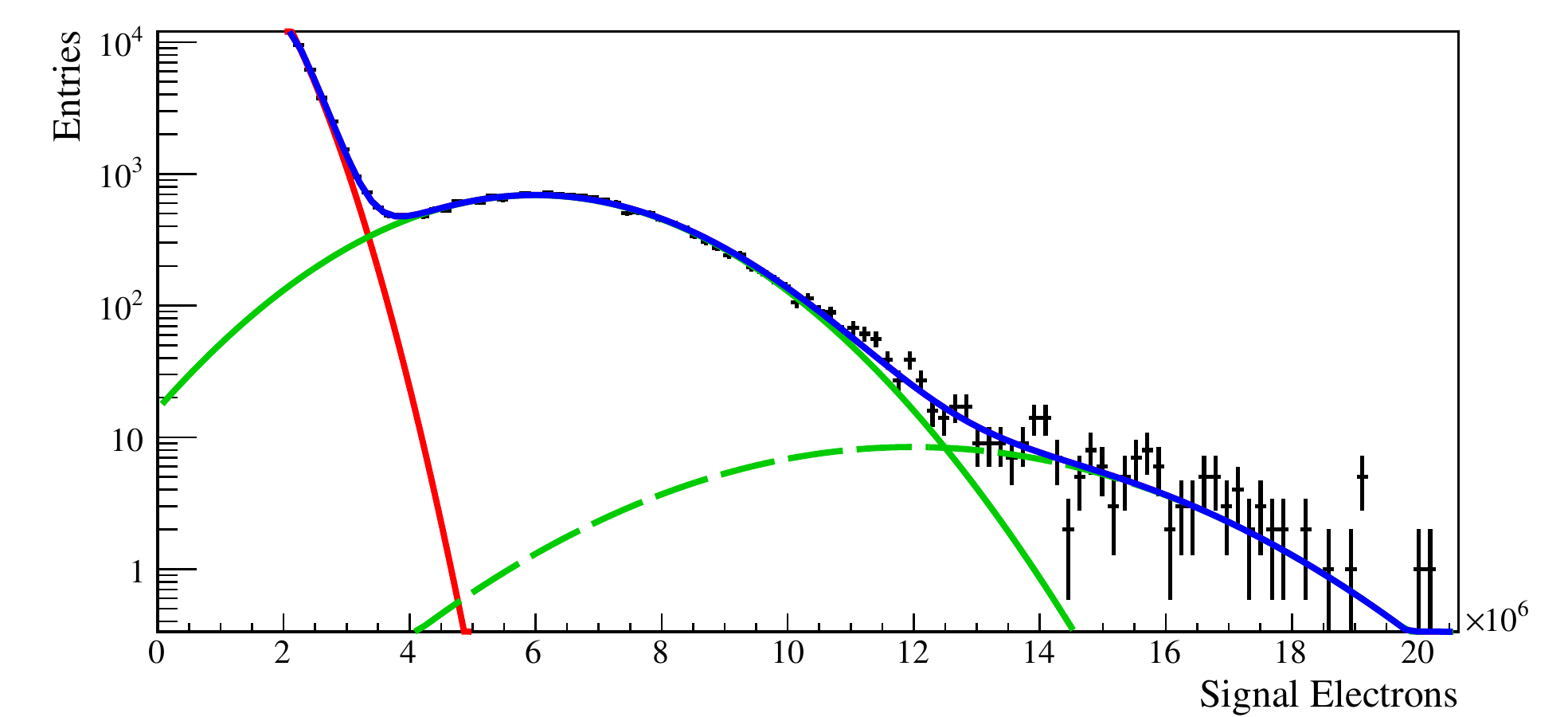}
\caption{SPE spectrum of a R11410 PMT immersed in LXe by stimulating the emission of SPEs using blue LED light. The lowest peak is due to noise (red), the main one from SPEs (solid green), and the tail from two PEs (dashed green). The spectrum is fit by a sum of Gaussians, Equation~(2.1), and used to derive the PMT gain from the position of the SPE peak. The full fit-function is shown in blue.}
\label{fig::spe}
\end{figure}

A peak processor scans the digitized waveforms for excursions from the baseline, integrates the area around the maximal excursion to obtain the number of electrons contributing to the signal, and histograms the result. One example is shown in Figure~\ref{fig::spe}. The resulting single photoelectron spectrum consists of a large peak around zero, which is due to electronic noise, followed by a (ideally well separated) peak from SPEs. In the analysis the spectrum is described by a sum of three Gaussians 
\begin{equation}\label{eq::fitfunc}
f(x) = a_n \times \textnormal{Gauss}(\mu_n,\sigma_n) + a_{\mathrm{SPE}} \times \textnormal{Gauss}(\mu_{\mathrm{SPE}},\sigma_{\mathrm{SPE}}) + a_{2} \times \textnormal{Gauss}(2 \mu_{\mathrm{SPE}}, \sqrt{2}\sigma_{\mathrm{SPE}})
\end{equation}
considering noise ($n$), the SPE peak, and the peak from two photoelectrons (2). $\mu_{\mathrm{SPE}}$ is given by the gain. $\sigma_{\mathrm{SPE}}$ is the SPE resolution. Mean and $\sigma$ of the Gaussian describing two photoelectrons (PEs) are fixed by the parameters of the SPE peak. Its amplitude $a_2$ is typically only 2-3\% of $a_{\mathrm{SPE}}$.

In some cases, the measurement was triggered by the PMT signal itself using a threshold discriminator (LeCroy 621S) to generate the trigger, which led to the same results.

\paragraph{Room-Temperature Characterization}

The basic characterization, as presented in Section~\ref{sec::character}, was performed at room temperature in a light-tight black box which can accommodate two R11410 PMTs fixed in a PTFE structure. Cables were guided into the box via a light-tight custom made feedthrough. In order to perform the measurements in a stable and reproducible configuration, they were started not earlier than 30\,minutes after switching the PMT on.

\paragraph{MarmotXS}

\begin{figure}[tb]
\centering
\includegraphics[width=0.95\columnwidth]{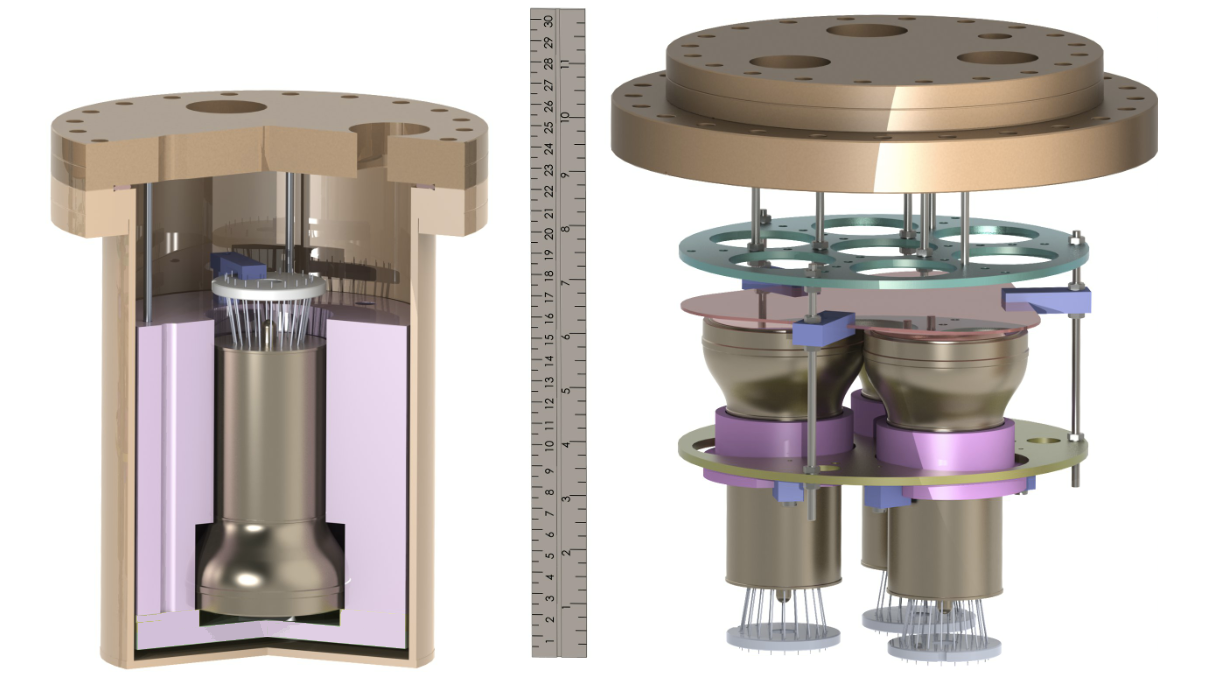}
\caption{Experimental setups for the performance tests in xenon. (Left) The small single phase chamber MarmotXS can accommodate one R11410 and was used to perform long-term stability measurements in LXe as well as cooling-cycling tests. (Right) MarmotXL housed three R11410 tubes, testing realistic arrangements in the xenon gas phase under various electric field configurations. Parts made of PTFE are drawn in magenta to enhance their visibility.}
\label{fig:SetupMarmot}
\end{figure}

To monitor the performance of the R11410 immersed in LXe over several months, a simple detector chamber, MarmotXS, was designed and built. The cylindrical chamber has an inner diameter of 15\,cm and a length of 23\,cm, see Figure~\ref{fig:SetupMarmot} (left) and allows the installation of one R11410 PMT facing downwards. To minimize the amount of LXe needed, most of the space is filled with a PTFE spacer surrounding the PMT which monitors a 5\,mm thick LXe layer below the photocathode. Typically, MarmotXS is filled with 400\,g of LXe, which covers about 2/3 of the PMT but leaves the back of the PMT and the HV~divider in the gas phase.

Cooling is achieved by placing the chamber in the gas phase above a bath of liquid nitrogen (LN$_2$) with the cooling power being uniformly distributed to the sides of the stainless steel chamber using a 3\,mm thick copper conductor. The single-wall cryostat is placed inside an open flask LN$_2$ dewar which is filled using an automated system. A PID control loop on one of the two temperature sensors in the setup provides the necessary stability and also supplies a maximum of 25\,W of heating power to the chamber walls. The xenon gas pressure and temperature are constantly monitored and are stable within $\pm0.2$\,bar and $\pm1$\,K, respectively. A blue LED is installed inside the PTFE filler to diffuse the light and to allow regular calibrations to monitor the PMT gain. 

MarmotXS is connected to a xenon gas system to fill and empty the chamber, described in detail in~\cite{ref::kr83m}. The detector was filled through a high temperature getter to remove electronegative contamination from the xenon. 

\paragraph{MarmotXL}

MarmotXL is a larger LXe chamber, its schematic picture is shown in Figure~\ref{fig:SetupMarmot} (right). It features a pulse tube refrigerator (cooling power 18\,W at 77\,K) for cooling which allows a very stable operation over long time intervals. The cryostat and the upper flange are large enough to fit either a smaller (15\,cm diameter) or a larger (25\,cm) inner detector vessel surrounded by an insulation vacuum. For the tests performed in the context of this study, namely the performance of the PMTs when operated in xenon gas in an array at various voltage configurations (see Section~\ref{sec::conclusion}), the larger inner chamber was used, which can accommodate three R11410. 

It was filled with xenon gas at room temperature at an operation pressure around 2\,bar absolute. The xenon was extracted from the bottom of the chamber and constantly purified by passing it through a high-temperature getter, before feeding it back to the top of the vessel. In this mode, the high voltage performance in ultra-pure xenon gas was studied in a similar configuration as when operated in the top array of a LXe detector. 

The PMTs were installed in a triangular pattern facing upwards, with the possibility to adjust the PMT-to-PMT distance. A polished stainless steel plate mounted above the PMT cathodes at a variable distance was biased with a high voltage to study the impact of strong electric fields close to the photocathode.

\section{Characterization of the Photomultiplier}\label{sec::character}

A short description of the R11410 PMT is given in this Section, followed by a summary of measurements of its intrinsic radioactivity using a low-background material screening facility. The Section concludes with the report of several general performance tests which were carried out at room temperature.

\subsection{Description of the R11410}

The R11410 is a photomultiplier of 3''~diameter produced by Hamamatsu~\cite{ref::hamamatsu}. It is specifically designed for low temperature operation (down to $-110^\circ$C) such as in LXe. The quantum efficiency (QE) of its bialkali photocathode (Hamamatsu LT version) has a maximum of typically 30\% and higher at 178\,nm~\cite{ref::lung2012}, fully covering the scintillation wavelength spectrum of xenon. The PMT window, made of synthetic silica, is transparent at this wavelength. The collection efficiency (CE) is about 90\%~\cite{ref::lung2012,ref::akerib2012}.

The 12-dynode PMT with a typical gain of $5\times10^6$ has a cobalt free Kovar metal body of 11.4\,cm length and a maximal (minimal) diameter of 7.8\,cm (5.3\,cm). The photocathode has a minimum effective diameter of 6.4\,cm. The typical voltage applied between anode and cathode is around 1500\,V, with the maximum being 1750\,V. The high magnetic permeability of the Kovar (a Fe-Ni-Co alloy) allows the operation of the PMT in the Earth's magnetic field without magnetic shielding~\cite{ref::lung2012}.

Following the initial version of the R11410, which had an intrinsic radioactivity too high for the next generation of dark matter searches~\cite{ref:xeScreening}, a low radioactivity version (R11410-MOD, later renamed to R11410-10~\cite{ref::hamamatsuComm}) has been manufactured by Hamamatsu. This version was extensively tested in the study presented here.

\subsection{Intrinsic Radioactivity}\label{sec::radioactivity}

Two of the R11410 tubes have been examined using {\it Gator}, a low-background, high-purity germanium detector~\cite{ref::gator} for $\gamma$-spectroscopy operated underground at LNGS by the Z\"{u}rich group. The results are summarized in Tab.~\ref{tab::screening}. For most isotopes, no significant evidence for a $\gamma$-line above the background could be found, leading to upper limits on the activity. The results from the two tubes, which came from from independent production batches, agree within their uncertainties. Compared to the much smaller R8520, a $2.54 \times 2.54$\,cm$^2$ PMT which is used in XENON100~\cite{ref::xeInstr, ref:xeScreening}, the intrinsic radioactivity of the tube is 3\,times smaller when normalized to unit active area, making it indeed a good candidate light sensor for next-generation experiments.

\begin{table}[h]
\centering
\caption{\label{tab::screening} HPGe screening results of two Hamamatsu R11410 PMTs using the {\it Gator} facility at LNGS. To estimate the neutron background from $(\alpha,n)$ reactions correctly, it is important to take into account a possible disequilibrium in the decay chains and to quote the activities for the different parts of the chains individually.} 
\begin{tabular}{l|c|ccccccccc}
 \hline \hline
& Livetime & Units & $^{238}$U & $^{226}$Ra & $^{228}$Ra & $^{228}$Th & $^{235}$U & $^{40}$K & $^{60}$Co & $^{137}$Cs \\
\hline  
\#1 & 17 days & mBq/PMT & $<$30 & $<$2.1 & $<$3.4 & $<$2.2 & $<$2.0 & 12$\pm$4 & 3.6$\pm$0.5 & $<$1.2 \\
\#2 & 33 days & mBq/PMT & $<$18 & $<$1.4 & $<$2.9 & $<$2.0 & $<$1.5 & 17$\pm$3 & 4.3$\pm$0.4 & $<$0.7 \\
\hline \hline
\end{tabular}
\end{table}

Two remarks are in order: First, the R11410 is continuously optimized in terms of radioactivity and a new version is under development. Second, for large scale LXe detectors which exploit the self-shielding capabilities of LXe by fiducialization, the most relevant background is from neutrons, which in case of the PMTs are produced via $(\alpha,n)$ and spontaneous fission reactions in the materials. The $\alpha$-particles are also produced in the early parts of the $^{238}$U and $^{232}$Th chains, which can be examined with superior sensitivity using methods such as mass spectrometry. These results will be published elsewhere.

\subsection{Room-Temperature Performance}

The general performance of the R11410~PMT at room-temperature is detailed in~\cite{ref::lung2012}. We summarize our own measurements here, the results agree with previous works. The resistor values of the voltage divider chain were 4\,:\,1.5\,:\,2\,:1\,:\,1\,:\,1\,:\,1\,:\,1\,:\,1\,:\,1\,:\,1\,:\,2\,:\,1\,M$\Omega$ between cathode, dynodes 1-12, and ground.

\begin{figure}[b!]
\centering
\includegraphics[width=0.49\columnwidth]{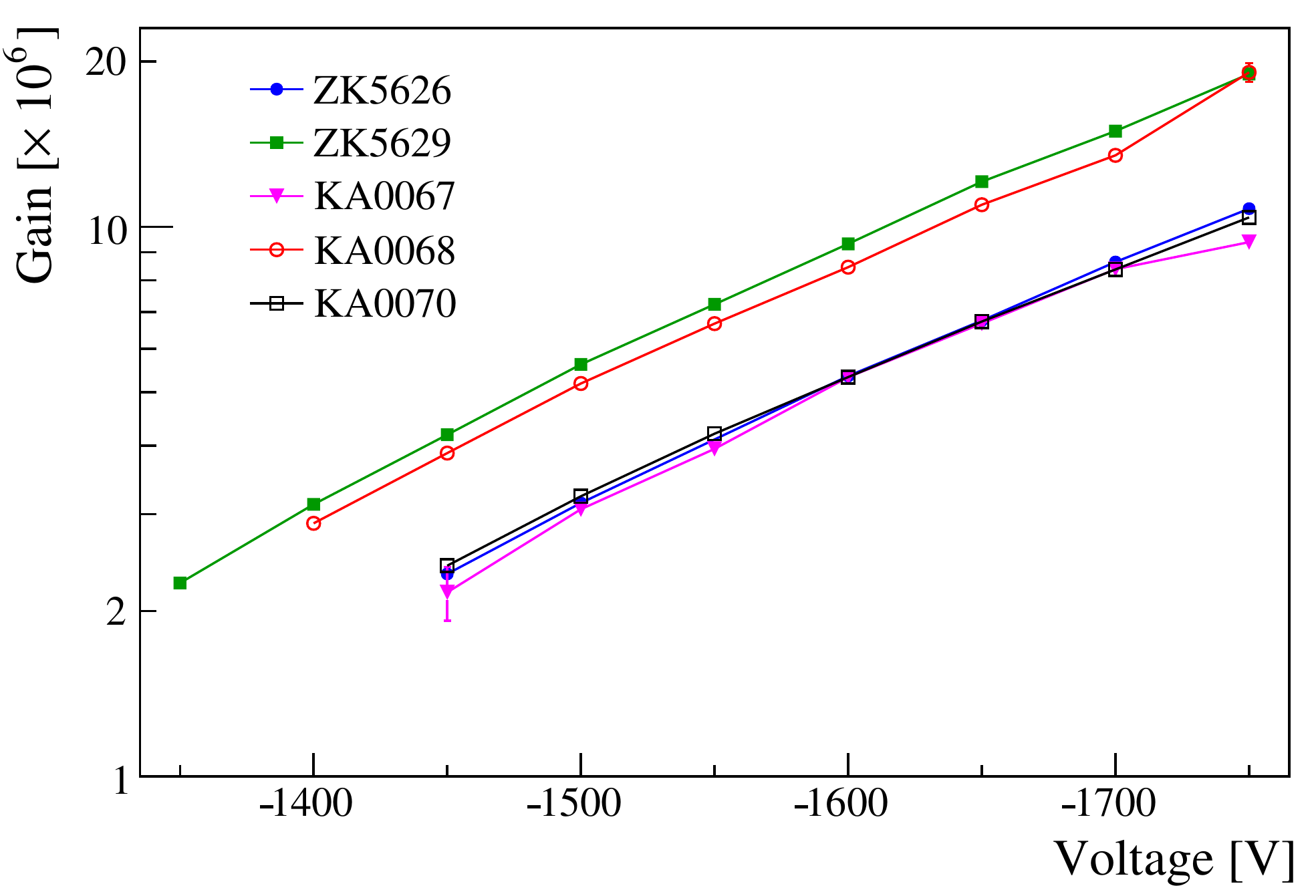} 
\includegraphics[width=0.49\columnwidth]{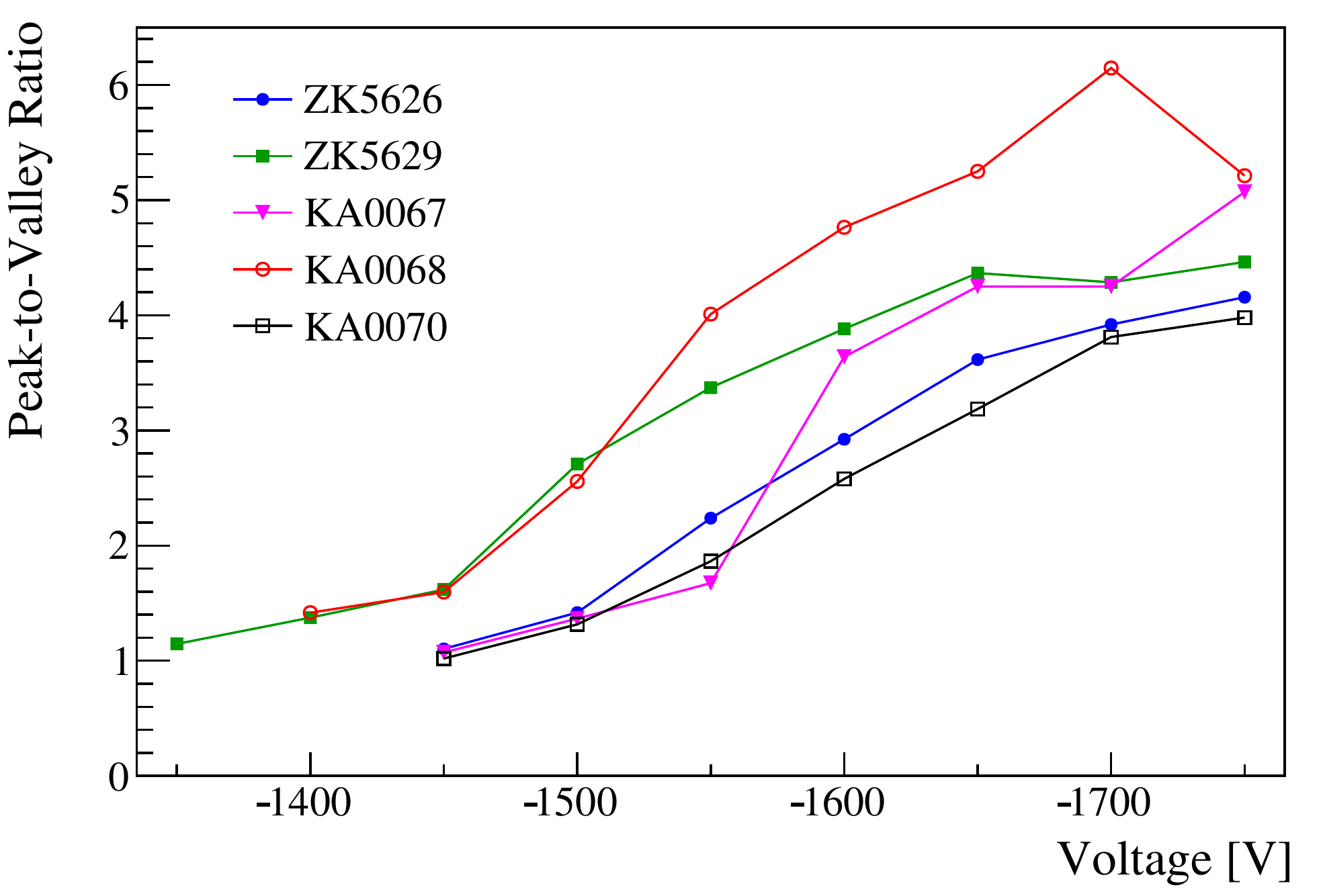} 
\caption{Gain (left) and peak-to-valley ratio (right) for five R11410 PMTs, measured vs.~negative high voltage. The tubes are from two different batches as indicated by the ZK and KA serial numbers.}
\label{fig:GainPtoV}
\end{figure}

\paragraph{Gain versus HV} 
According to Hamamatsu's specifications, the typical gain of the R11410 is  $\sim 5 \times 10^6$ at a bias voltage of $-1500$\,V. In the black box we have measured the gain of five PMTs at different voltages from $-1350$\,V to $-1750$\,V in 50\,V steps. Not in all cases the SPE peak could be well separated from the noise at the minimum voltage. The results, shown in Figure~\ref{fig:GainPtoV} (left), indicate that there is a $\pm20$\% variation around the typical gain. The expected exponential dependence on the bias voltage is observed in all cases. 

The SPE resolution of the PMTs is given by the ratio $\sigma_{\mathrm{SPE}}/\mu_{\mathrm{SPE}}$ of the parameters of Equation~(\ref{eq::fitfunc}). It is almost independent of the applied high voltage and varies between 35\% and 40\% for all tested units.

\paragraph{Peak-to-Valley Ratio}
The ability to separate the SPE spectrum from the electronic baseline noise can be quantified by the peak-to-valley ratio, i.e.~the maximum of the SPE peak divided by the height of the valley which separates it from the noise. Maximum and valley are directly taken from the measured spectrum and the peak-to-ratio is shown in Figure~\ref{fig:GainPtoV} (right): it slightly increases with the bias voltage, and very good values of~3 and above are reached at the nominal voltage of $-1500$\,V for two of the five tested PMTs. The ratio is $\sim$15\% lower for the remaining tubes but at a bias voltage of $-1600$\,V, all PMTs have a peak-to-valley ratio above~3. This ratio was only achieved by a small minority of tubes used successfully in XENON100~\cite{ref::xeInstr}. The PMTs with a higher gain also tend to have a higher peak-to-valley ratio. 

\paragraph{Afterpulses} 
Residual-gas molecules inside the PMT are either absorbed by its getter or are trapped in the surfaces. Molecules trapped by the first dynode can be ionized by light-induced photoelectrons and then drift to the PMT cathode where they release further electrons. The amplitude of these delayed afterpulses varies depending on the type of residual ion and on the voltage applied to the PMT. Afterpulses can affect the accurate measurement of small few-photoelectron signals following a large amplitude pulse and thus must be studied and understood.

\begin{figure}[!h]
\centering
\includegraphics[width=0.48\columnwidth]{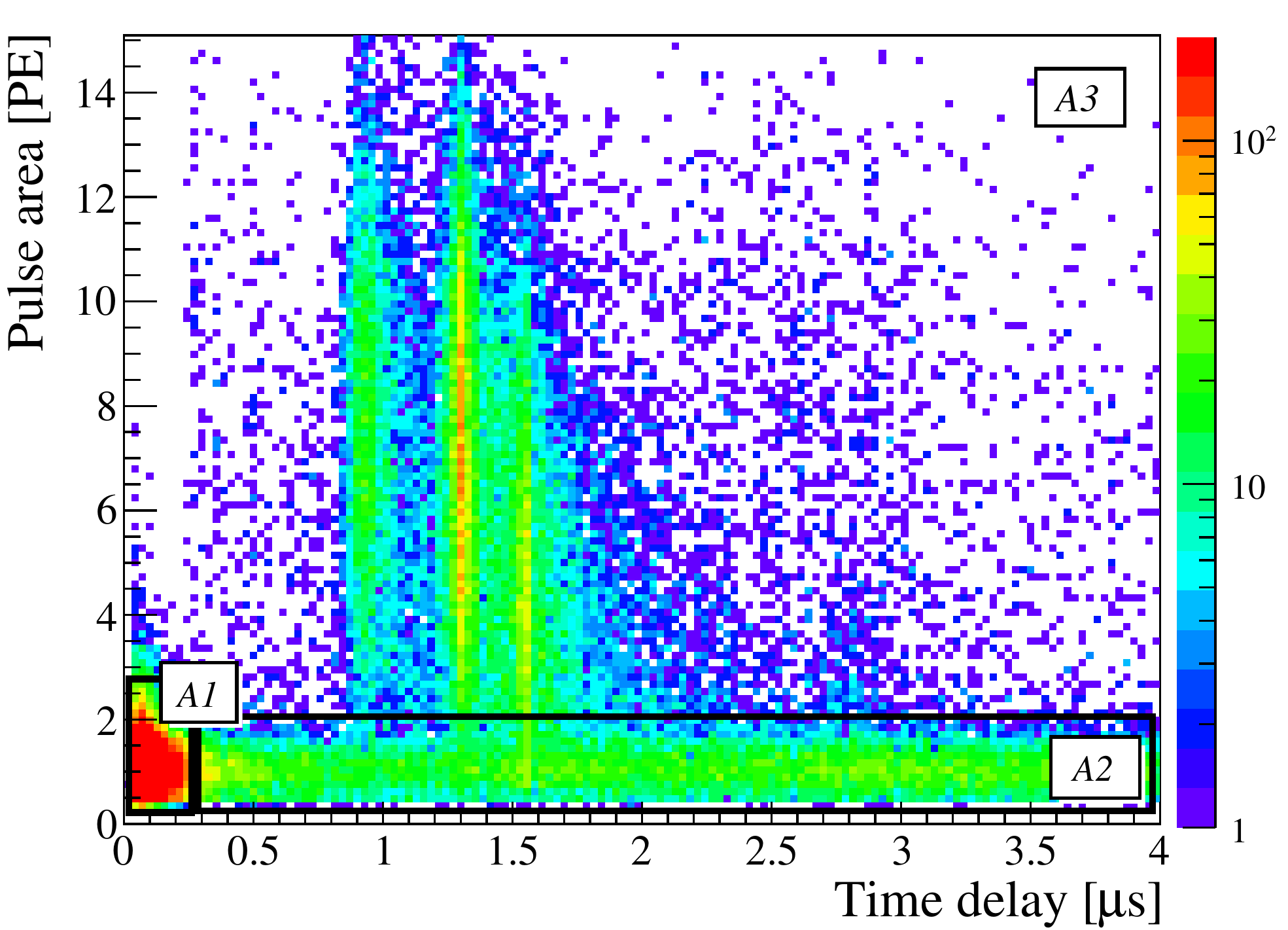}
\includegraphics[width=0.49\columnwidth]{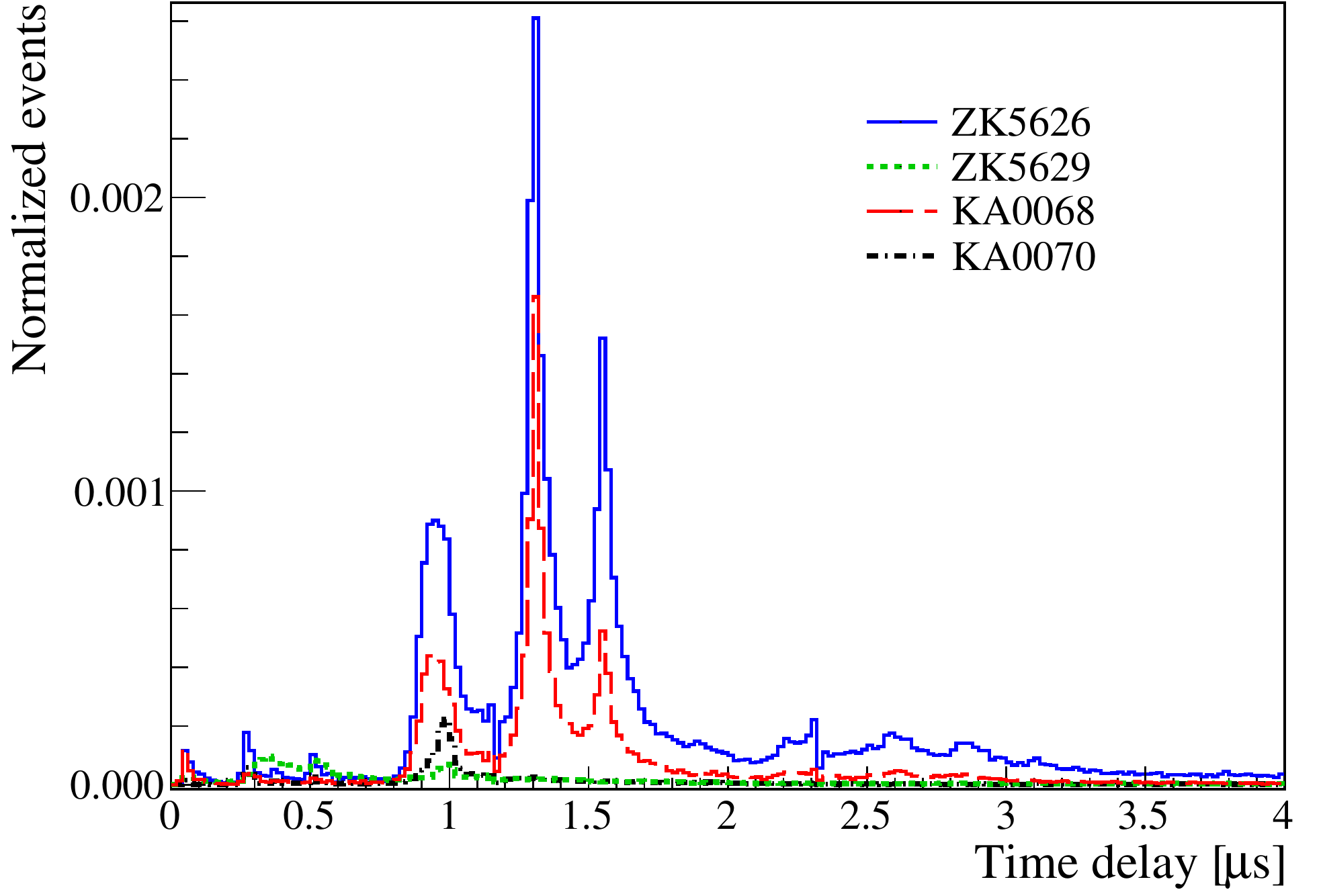}
\caption{(Left) Afterpulse area vs.~time spectrum for an R11410~PMT (KA0068). Three regions can be identified, with clearly visible lines extending up to several PE in region A3. (Right) Time spectra of region A3. The peaks stem from specific ionized molecules, their relative size and abundance vary for each PMT. }
\label{fig::afterp}
\end{figure}

We have studied afterpulses by externally triggering on dark current pulses and searching for peaks following the triggering signal. As an example, Figure~\ref{fig::afterp} (left) shows the pulse area of an R11410 PMT (KA0068) measured vs.~the time after a SPE event. Three distinct regions have been identified according to the characteristics of the afterpulses.
A1: within a delay of several tens of nanoseconds, a large number of afterpulses are measured with a signal mean of about 1\,PE. 
A2: afterpulses of 1\,PE are measured continuously at all times.  
A3: at distinct time positions, afterpulse peaks appear with pulse areas of up to tens of PE, see also Figure~\ref{fig::afterp} (right) where the A3 spectra of several PMTs are shown normalized to the total number of events. The peaks are due to specific ionized molecules which are liberated from the first dynode by the initial photoelectrons. Depending on mass and ionization level, the geometry and field in the PMT leads to individual peaks as in a mass spectrometer. Following~\cite{ref::lung2012}, the peaks at 0.3\,$\mu$s and 1.0\,$\mu$s can be associated with H$^+_2$ and CH$^+_4$, respectively. The large peaks around 1.3\,$\mu$s and 1.6\,$\mu$s, which are only present in two PMTs, could be due to CO$^+$ and CO$_2^+$~molecules. 

The total afterpulse rates, as well as the contribution of each region in Figure~\ref{fig::afterp} (left), is given in Tab.~\ref{tab::afterpulses}, along with the specific contributions of the peaks observed at various times. We observe rather large differences between the 5~studied PMTs, with the total afterpulse rate differing by up to a factor~10. We did not find a correlation with the PMT production batch. 

\begin{table}[h]
\centering
\caption{\label{tab::afterpulses} Afterpulse rate as a fraction [\%] of the total number of acquired events. The total rate (fraction of events followed by an afterpulse) is given, as well as the contributions of the individual populations (A1, A2, A3) and the most dominant peaks (at specific delay times) identified in Figs.~4. The statistical uncertainties are small, the systematic uncertainty is estimated to be about $\pm$5\% as determined by repeating the measurements 3\,times. } 
\begin{tabular}{l|c|ccc|ccccc}
 \hline \hline
PMT & Total \% &  A1  &  A2  &  A3  &  0.27\,$\mu$s &  0.52\,$\mu$s & 0.98\,$\mu$s &  1.32\,$\mu$s &  1.56\,$\mu$s \\ \hline
ZK5626 & 8.6 & 3.4 & 2.0 & 3.2 & 0.04 & 0.03 & 0.62 & 0.99 & 0.66 \\
ZK5629 & 2.2 & 0.5 & 1.5 & 0.3 & 0.04 & 0.07 & 0.05 & 0.02 & 0.01  \\
KA0068 & 4.6 & 2.3 & 0.9 & 1.4 & 0.02 & 0.01 & 0.29 & 0.55 & 0.24  \\
KA0067 & 1.3 & 0.1 & 0.8 & 0.5 & 0.13 & 0.12 & 0.04 & 0.01 & 0.01  \\
KA0070 & 1.1 & 0.3 & 0.6 & 0.2 & 0.01 & 0.01 & 0.11 & 0.02 & 0.01  \\
\hline \hline
\end{tabular} 
\end{table}

For one PMT (KA0067) the afterpulse rate was also studied at LXe temperatures, right after xenon recuperation and subsequent evacuation (see Section~\ref{sec::cycling}). The rate was about~20\% lower than the one observed at room temperature. This is expected since some of the rest gas which is responsible for the afterpulses gets attached to the cold PMT inner surface.

Afterpulses at the level of a few percent do not significantly affect the capability of LXe detectors to operate in dark matter search mode. In recent analyses of LXe detectors~\cite{ref::xe100_analysis}, an event would be discarded if two candidate light pulses are observed in a waveform. In order to reduce the impact of dark counts, a candidate pulse has to be observed by at least 2~PMTs at the same time. The low energy signals of interest are typically seen by less than 15~PMTs and the likelihood that two of these tubes see an afterpulse (assuming an average rate of 3\%) within $\Delta t = 30$\,ns is only $\sim 0.15$\% and leads to a negligible acceptance loss.

\section{Long-Term Stability Tests in Liquid Xenon}\label{sec::lxestability}

The R11410 was developed to be operated at cryogenic temperatures, however, it is mandatory to test its performance immersed in LXe before it can be considered as light sensor for a dark matter detector. We have performed several tests using the small single phase chamber MarmotXS, with the main focus on two aspects: the performance of the tube when operated continuously in LXe over several months, and the effect of rather abrupt temperature changes of $\Delta T \approx 120$\,K which arise when a LXe detector is filled or emptied.

\begin{figure}[!h]
\centering
\includegraphics[width=0.38\columnwidth]{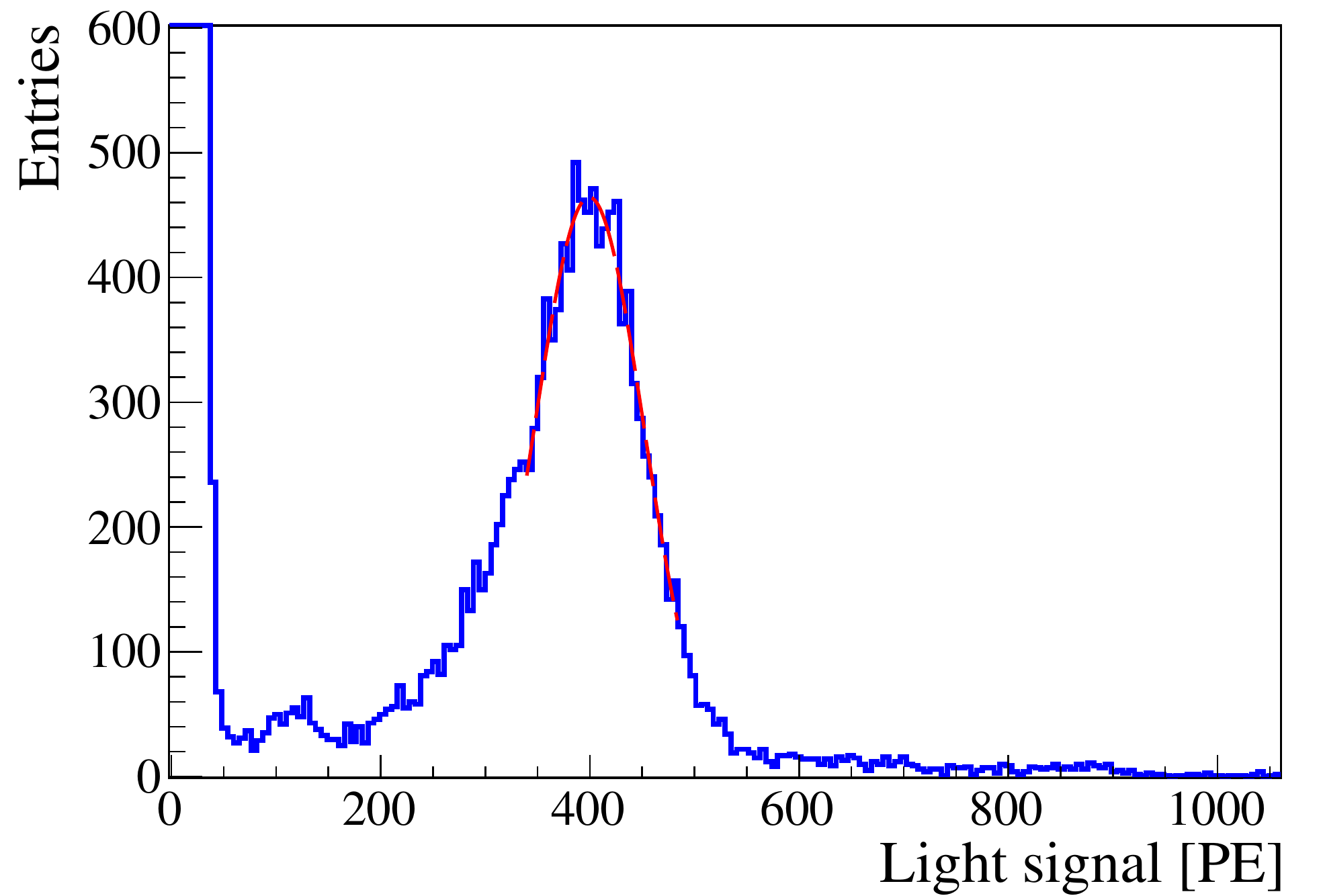}
\includegraphics[width=0.61\columnwidth]{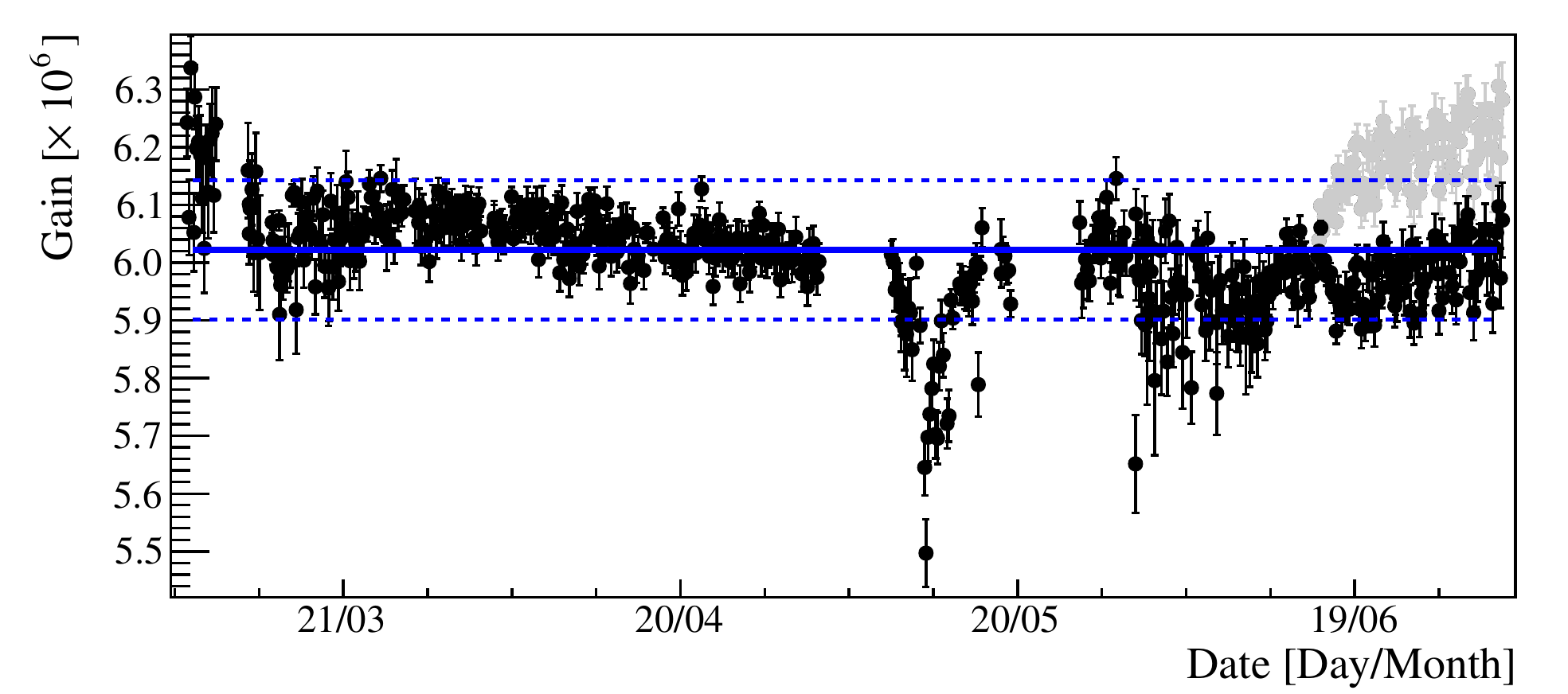}
\caption{(Left) $^{57}$Co scintillation light spectrum obtained with a R11410 immersed in LXe in MarmotXS. (Right) Long-term stability of the gain of a R11410 immersed in LXe, as measured over a period of about 5\,months. The gain is stable within $\pm2$\%, as indicated by the dashed lines. Periods in which the gain shows larger variations can be correlated to changing experimental conditions. At the end of the run, the gray points show the measurements while the corresponding black points are corrected and take into account a sudden change in the LXe pressure and temperature. A similar correction cannot be done reliably for the short spikes in the gain measurements, hence they are presented as measured.}
\label{fig:57CoLXe}
\end{figure}

The test setup was such that the R11410 was almost completely covered by LXe, with only the back part with the high voltage divider being in xenon gas. This is an extreme test condition as stronger temperature gradients will occur than in a dark matter experiment. Moreover, the PMT voltage divider is tested in the gas, which is a worse electrical insulator than LXe. MarmotXS was designed for these stability tests but not optimized for measurements using LXe as scintillator. Nevertheless, data were acquired with $\gamma$-calibration sources. The full absorption peak of $^{57}$Co exhibits a light yield of about 3.3\,PE/keV, as shown in Figure~\ref{fig:57CoLXe} (left) with the fit to the peak.

Overall, we have tested 3~tubes in LXe. One of them (the first version without the reduced radioactivity) was cooled down 3\,times, with the longest cold period being 6\,days. No definitive conclusions could be made because of instabilities of the cooling system. Longer LXe tests of two R11410, modified for a lower intrinsic radioactivity, using a more stable setup are described in detail below. 

\subsection{Gain Stability}

The gain of an R11410 has been measured over a period of almost 5\,months in which the PMT was operated without any relevant interruption at a bias voltage of $-$1600\,V. It was automatically measured every 4\,hours, with the exception of a few periods where the automatic system failed. Pressure and temperature of the xenon inside the chamber were constantly monitored. As shown in Figure~\ref{fig:57CoLXe} (right) the gain was found to be stable within $\pm$2\%, which is the intrinsic uncertainty of the method to derive the gain~\cite{ref::xeInstr} by fitting the function given by Equation~(\ref{eq::fitfunc}) to the SPE spectrum. 

At a few occasions, the gain response deviates slightly from this stable behavior, however, these periods can be related to changing experimental conditions: At the beginning of the long-term test, the LXe system was not yet in a stable state and the thermodynamic conditions were still changing. The two deviations to lower gain values are most likely not caused by a real gain change, but by a change in the baseline noise condition which has a small but significant impact on the gain determination. In one case, the voltage of the LED to stimulate SPE emission was set too high and too much light illuminated the PMT for a few hours. In the other case, a malfunction in the cooling system lead to a sudden temperature decrease followed by an increased noise level. After all occasions, the stable gain value was reached again. The continuous gain increase observed at the end of the measurement (gray points) is directly correlated with a slight increase in the temperature ($+1$\,K) and pressure ($+100$\,mbar) inside the chamber. The corresponding black points have been corrected for this effect by subtracting the difference of the average gains before and after the temperature/pressure increase.

Following this successful long-term measurement, another R11410 was operated in LXe with the focus on its response to thermal stress associated with repeated cooling cycles. The measurements are reported in the next Section.

\subsection{Stress Tests with Cooling Cycles}\label{sec::cycling}

When dark matter detectors operated with LXe are filled, the setup is being cooled down from room temperature to about $-100^\circ$C in rather short time of a few hours which imposes thermal stress on the light sensors. In the past, we experienced that a small fraction of PMTs (XENON100 1-2\%~\cite{ref::xeInstr}, using a different PMT model) were not operational after the first cool-down of the detector. In order to test how the R11410 behaves when exposed to thermal stress, we subsequently went through several cooling cycles with two PMTs. (A third PMT has been only cooled down twice.) After a cool-down, the temperature was kept stable at $(-99 \pm 1)^\circ$C for typically 5\,days, then raised and kept at $(21 \pm 3)^\circ$C for about 5\,days, before the chamber was filled with LXe again. The PMTs were not powered during cool-down or warm-up. None of the tested PMTs showed any indication of a worsened response or malfunctioning during these tests. However, we emphasize that due to the limited number of considered PMTs, the tests reported here cannot identify cool-down problems which might occur at the percent level. Therefore, every tube which is to be installed in a LXe detector should be cooled-down at least once before detector assembly.

\begin{figure}[!h]
\centering
\includegraphics[width=0.99\columnwidth]{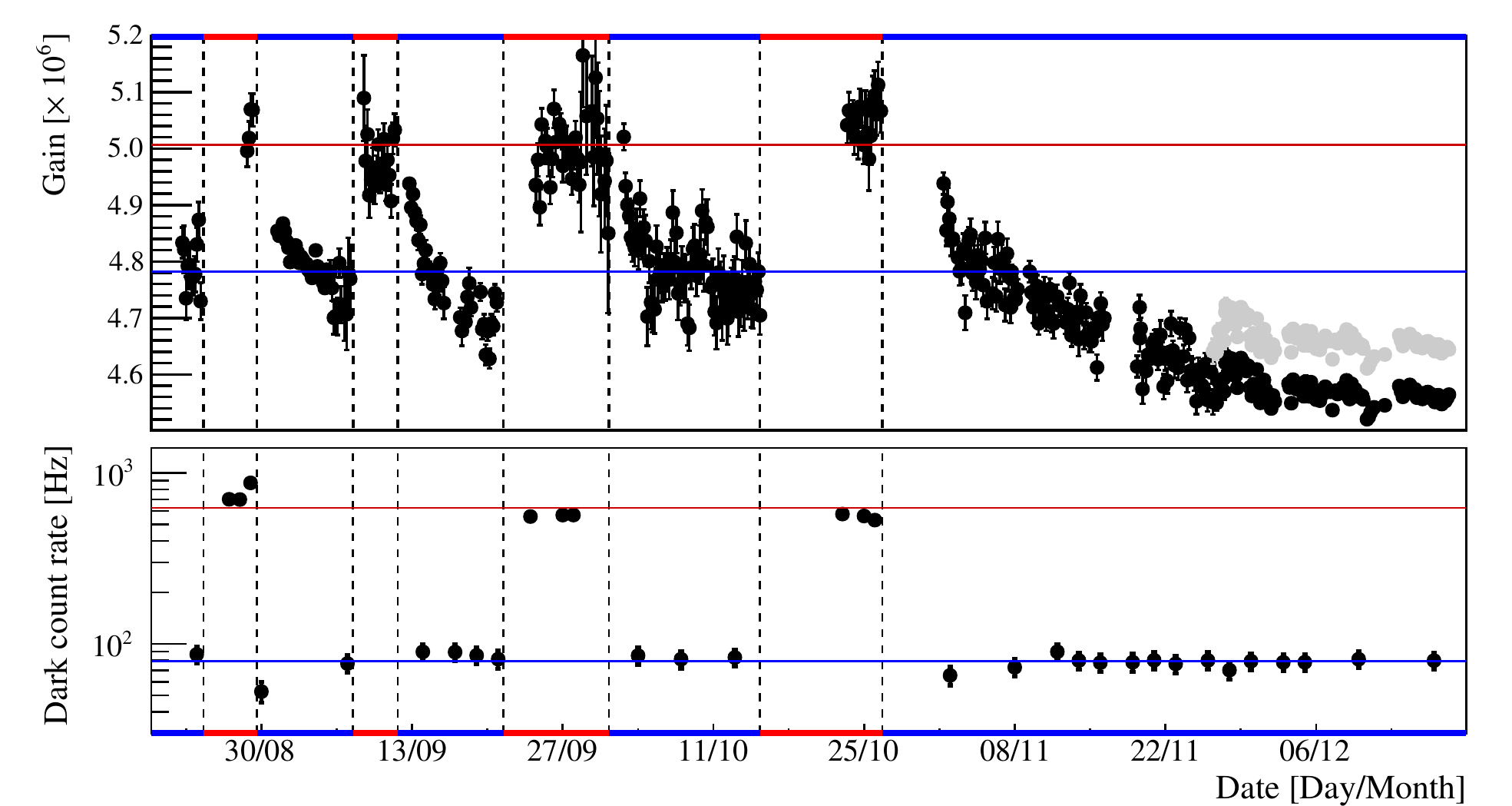}
\caption{Time evolution of the gain (top panel) and dark count rate (bottom) of a R11410 PMT during thermal cycling. The tube was immersed in LXe 5\,times and warmed up in between, with each period lasting from 5\,days to several weeks. After cool-down the gain decreased by $\sim$5\%, while the dark count rate was reduced by an order of magnitude. The dark count rate shown here was measured in xenon environments and is therefore an upper limit as real scintillation pulses are also counted. The corrected rate at LXe temperatures is a factor~2 lower (see text). At the end of the run the gain measurements were affected by a changing xenon temperature/pressure: The measured values are shown in gray, the ones corrected for this effect are shown in black. The dark count measurements are not corrected. }
\label{fig:CoolingCycles}
\end{figure}

Figure~\ref{fig:CoolingCycles} (top) shows the gain evolution during 5\,cooling cycles performed over more than 3\,months. The gain of a PMT biased at $-1600$\,V was measured using a blue LED at room temperature and at LXe temperature, marked with red and blue horizontal bars, respectively. The gain at room temperature was systematically and reproducibly found to be about 5\% higher than at LXe temperature. However, it took more than 4\,weeks to reach a stable gain after cool-down at a $\sim$10\% lower level (compare to Figure~\ref{fig:57CoLXe}, right).

Additionally, the dark count rate of the PMT was measured one hour after the LED for the gain measurement was turned off. The PMT signal was amplified and data were taken triggering by a discriminator. Figure\,\ref{fig:CoolingCycles} (bottom) shows the evolution of the dark count rate defined by the integral of the Gaussian function describing the SPE peak, the second part of Equation~(\ref{eq::fitfunc}). 
The average dark count rate at room temperature was ($630\pm110$)\,Hz and ($79\pm8$)\,Hz at LXe temperature. The rate was determined by integrating the SPE peak above 0.3\,PE, and the uncertainties are given by the standard deviation of all measurements. It was stable during the whole measurement period. Note that the chamber was filled with gas or liquid xenon at room and LXe temperature, respectively. Therefore these values represent upper limits on the actual dark count rate, since scintillation light is also observed.

To avoid bias by this effect, the dark count rate was also measured in vacuum at the end of the temperature cycling immediately after LXe recuperation, when the temperature inside the chamber was still $-73^\circ$C, about $30^\circ$C warmer than the temperature when filled with LXe. With ($40 \pm 8)$\,Hz, the measured dark count rate was by about a factor of two lower than in cold xenon gas. This value increased to ($415 \pm 20$)\,Hz when the PMT was operated at room temperature (and in vacuum).

The accidental time coincidence of two dark count pulses can lead to fake scintillation light pulse candidates. If this happens within an event where the real light and charge signal are clearly visible, the event would be discarded at analysis level~\cite{ref::xe100_analysis}. Only cases where the real light signal is not found in the trace, the accidental signal could be falsely identified, potentially leading to fake WIMP-like events. Since various parameters must accidentally fall in the right range (e.g., (false) light to charge ratio, (false) interaction depth, etc.) this scenario is not very likely, but has to be carefully evaluated for every large dark matter experiment. At a dark count rate of 40\,Hz, an accidental rate of $\sim$1.5\,Hz is expected from 250\,PMTs, the typical number of channels for a ton-scale LXe detector~\cite{ref::xe1t} and assuming a time coincidence window of 30\,ns~\cite{ref::xe100_analysis}. While this value cannot be directly translated into an expected number of background events (this analysis requires more detector specific information), it is clear that a smaller dark count rate is beneficial. From this point of view, the low rate of 40\,Hz as measured here is an important feature of the R11410.

\section{High-voltage Tests in Gaseous Xenon}\label{sec::hvstability}

In dual-phase TPCs, about half of the PMTs are operated immersed in LXe below the target region. The other half are located in the gas phase above in order to detect the localized proportional scintillation light, providing information on the location of the event. (Note that in ZEPLIN-III~\cite{ref::zeplin} and the proposed design of the DarkSide-50 liquid argon detector~\cite{ref::darkside}, all PMTs are immersed in the liquid.) While LXe is a very good electrical insulator, problems related to the PMT bias voltage might arise in the gas phase. This is especially important when a negative voltage is applied to the PMT cathode which means that the whole metal case of the R11410 is at high voltage. However, this is the preferred operation mode for using the PMT in a LXe dark matter detector as it reduces the radioactivity since a HV-decoupling capacitor is not needed, and it also avoids possible ripples on the signals which are induced by the high voltage biasing the PMT.

\subsection{Tests in a Realistic PMT Arrangement}

In order to find out the minimal required distance between PMTs, which is relevant for the design of large scale PMT arrays, and to study the performance of such an array in different voltage configurations, a test setup of 3~R11410 PMTs was built and installed in MarmotXL. It features a variable distance between the PMTs and has an additional HV electrode (''plate``) installed parallel to the photocathode to test the PMTs in an external electric field. Various voltage configurations of the 3~tubes and the plate have been tried, mimicking situations which can occur in a large PMT array. While the typical bias voltage is around $-$1500\,V, some tubes might be set to higher/lower voltages to equalize the gain in the array, or a malfunctioning PMT might even require to be switched off. The latter would lead to a rather strong field between working and non-working PMTs ($\sim 5$\,kV/cm).

Allowing for some space for a support structure, the smallest anticipated center-to-center distance between the 77.5\,mm diameter tubes is 80\,mm. All configurations of PMT voltages tested at this distance are summarized in Tab.~\ref{tab::hvtest}. The xenon gas pressure (temperature) was 2.1\,bar absolute ($28^\circ$C). The gas was continuously purified at a rate of 10\,standard liters per minute. In all configurations the voltage settings were kept for at least one hour, the PMT currents were monitored, and the trip-current were set such that a small discharge would switch off the channel. Moreover, the PMT traces were checked occasionally. At 80\,mm minimal distance between the PMTs, no high voltage related problems were encountered. 

\begin{table}[h]
\centering
\caption{\label{tab::hvtest} Realized high-voltage configurations in the 3~PMT test array operated in high-purity xenon gas. The central distance between the PMTs was 80\,mm, and 15\,mm to the electrode plate. No high voltage related problem was experienced in any configuration. } 
\begin{tabular}{l|cccc}
 \hline \hline
& PMT~1 [V] & PMT~2 [V] & PMT~3 [V] & HV~plate [V] \\
\hline  
\#1 & $-$1600\,V & $-$1600\,V & $-$1600\,V & 0\,V \\
\#2 & $-$1750\,V & $-$1600\,V & $-$1500\,V & 0\,V \\
\#3 & 0\,V    & $-$1750\,V & $-$1750\,V & 0\,V \\
\#4 & $-$1750\,V & $-$1750\,V & $-$0\,V    & 0\,V \\
\#5 & $-$1750\,V & $-$1745\,V & $-$0\,V    & $+$1500\,V \\
\#6 & 0\,V    & $-$1750\,V & $-$1750\,V & $+$1500\,V \\
\hline \hline
\end{tabular}
\end{table}

\subsection{Performance in the Vicinity of a Strong Electric Field}

The PMTs installed below the TPC in a dark matter detector have to operate close to the TPC's cathode electrode. Often a screening grid is placed between PMT and electrode in order to shield the PMT from the strong electric field. This grid, however, absorbs some scintillation light and, because of its typically thin wires, might also induce high voltage related problems. It typically features a rather large geometric opening in order to reduce light absorption, which in turn allows for larger electric field reach-throughs. Hence it is mandatory to ensure that the PMT is operational with undisturbed performance close to relatively high electric fields. Possible problems could come from charge build up in the quartz window creating light signals in sudden discharges.

\begin{figure}[!h]
\centering
\includegraphics[width=0.8\columnwidth]{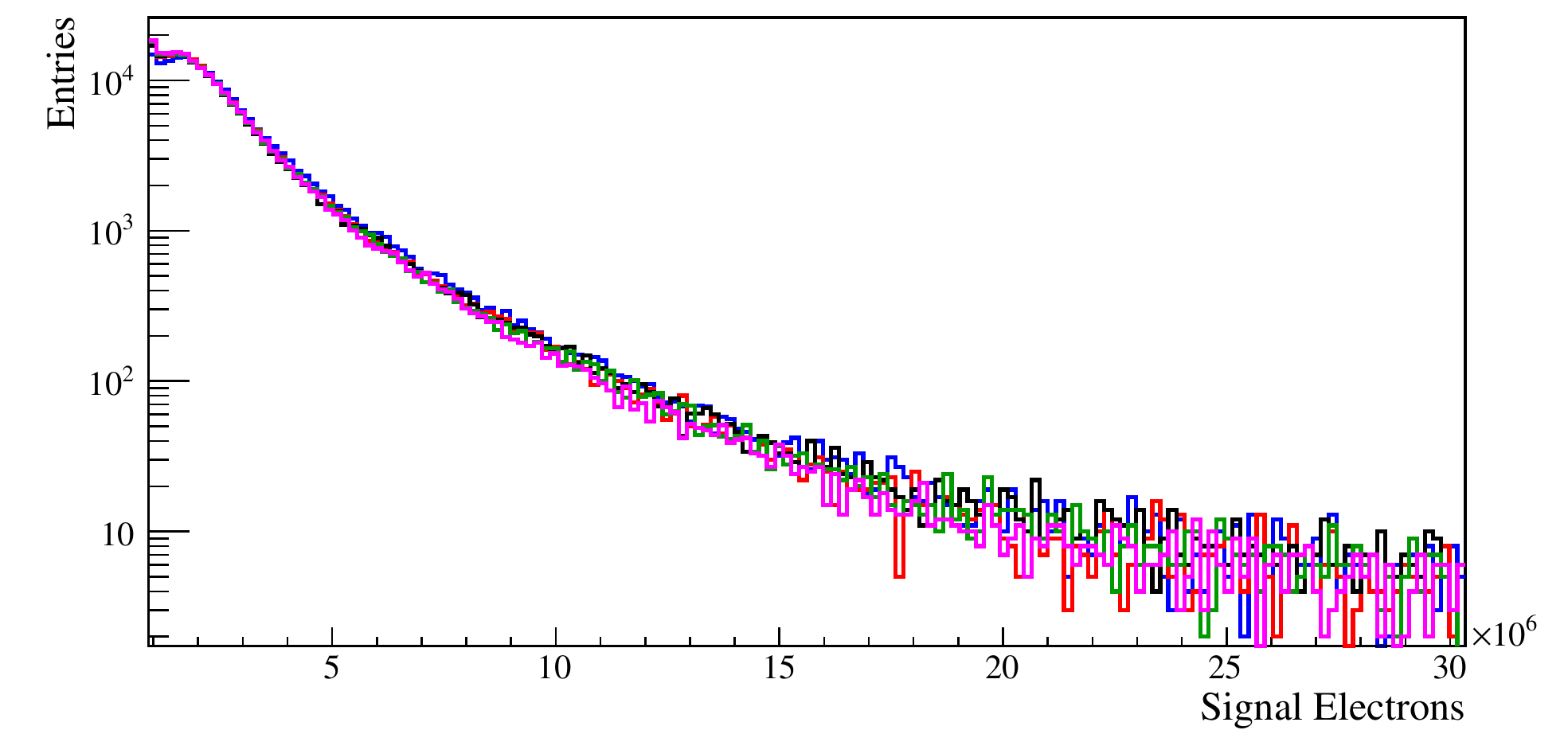}
\caption{Single photoelectron spectra of a R11410 PMT, operated close to electric fields of 0\,kV/cm, 7.5\,kV/cm, 10\,kV/cm, and 11\,kV/cm as indicated by the different colors. The $x$-axis is extended to $>$10\,PE in order to also observe larger light signals possibly induced by the field. No significant field dependence is detected, the PMT spectra are unaffected by the field above its photocathode.}
\label{fig:HVtest}
\end{figure}

This has been tested with the same three~R11410 PMT array by moving the electrode plate very close to the PMT cathode while biasing it with a positive high voltage. PMT cathode and metal case have been kept at $-$1500\,V, thus generating fields of 5\,kV/cm, 7.5\,kV/cm, 10\,kV/cm, and 11\,kV/cm when biasing the plate with 0\,kV, +0.75\,kV, +1.5\,kV, and +1.8\,kV, respectively. All PMTs remained fully operational, and no variation in their response could be measured. This is shown in Figure~\ref{fig:HVtest} which does not indicate any excess of events up to $>10$\,PE. We conclude that the R11410 can be operated in fields up to 11\,kV/cm. Higher fields might be possible but could not be tested in our setup due to voltage instabilities related to the high voltage on the electrode plate. When the R11410s are operated behind a screening grid, the expected remaining field strengths are well below the tested levels.

\section{Discussion and Conclusion}\label{sec::conclusion}

We have performed various tests with the Hamamatsu R11410-10 PMT, in setups which are relevant for the next-generation dark matter search experiments using LXe, examining the tube in configurations as they occur in a TPC. In particular, we have tested several units immersed in LXe and, using a different setup, in strong electric fields. We show that the R11410 features a stable gain while being operated continuously in LXe for several months. Repeated cooling cycles from room temperature to LXe temperatures around $-100^\circ$C were performed successfully without damaging the tested PMTs and without observing a change in their response. The PMTs successfully passed several high voltage tests in xenon gas, aiming to determine whether they could be operated next to each other at a minimal distance of 80\,mm and in strong electric fields. 

Additionally, we have characterized the PMT in terms of room temperature performance, and could largely confirm the results from Hamamatsu and Ref.~\cite{ref::lung2012}. The R11410 features a remarkable single photoelectron response with an excellent peak-to-valley ratio $\geq 3$ and a high gain around $5 \times 10^6$. We have observed that the afterpulse rate varies considerably between tubes (up to a factor~10), and can reach levels up to 10\% at room temperature. At LXe temperature, the afterpulse rate is slightly reduced by $\sim$20\%. At this temperature, the tube exhibits a very low dark count rate of $<50$\,Hz.

Sensitive photosensors with a low intrinsic radioactivity are mandatory for future dark matter detectors using liquid noble gases, such as xenon or argon, as WIMP targets. Based on the study presented here, we conclude that the Hamamatsu R11410 3''~PMT is a good candidate for LXe detectors. Therefore it was chosen to be the light detector for the next-generation experiment XENON1T~\cite{ref::xe1t}.

\acknowledgments
This work has been supported by the Swiss National Science Foundation (grant~\#200020\_138225) and by the FP7 Marie Curie-ITN action PITN-GA-2011-289442 {\it INVISIBLES}. We thank Hamamatsu for kindly providing several samples for these tests and M.~Metzger (Hamamatsu Switzerland) for his support. One early R11410 was kindly provided by K.~Ni (SJTU) for screening. We thank A.~James and D.~Florin (UZH) for their technical contributions and K.~Arisaka (UCLA) and E.~Aprile (Columbia) for many helpful discussions.

\end{document}